\documentclass[twocolumn]{aastex631}

\usepackage{graphicx}
\usepackage{amssymb}
\usepackage[T1]{fontenc}
\usepackage{lmodern}

\received{}
\revised{}
\accepted{}

\shorttitle{The KMTNet View of BLAPs}
\shortauthors{Kim et al.}

\begin{document}
\title{KMTNet View of Blue Large-amplitude Pulsators Toward the Galactic Bulge: \\ I. Discovery of Wide-orbit Companions in OGLE-BLAP-006}
\correspondingauthor{Seung-Lee Kim}
\email{slkim@kasi.re.kr}
\author[0000-0003-0562-5643]{Seung-Lee Kim}
\affil{Korea Astronomy and Space Science Institute, Daejeon 34055, Republic of Korea}

\author[0000-0003-0043-3925]{Chung-Uk Lee}
\affil{Korea Astronomy and Space Science Institute, Daejeon 34055, Republic of Korea}

\author[0000-0002-8692-2588]{Kyeongsoo Hong}
\affil{Korea Astronomy and Space Science Institute, Daejeon 34055, Republic of Korea}

\author[0000-0002-5739-9804]{Jae Woo Lee}
\affil{Korea Astronomy and Space Science Institute, Daejeon 34055, Republic of Korea}

\author[0000-0002-4292-9649]{Dong-Jin Kim}
\affil{Korea Astronomy and Space Science Institute, Daejeon 34055, Republic of Korea}

\author[0000-0002-7511-2950]{Sang-Mok Cha}
\affil{Korea Astronomy and Space Science Institute, Daejeon 34055, Republic of Korea}
\affil{School of Space Research, Kyung Hee University, Yongin 17104, Republic of Korea}

\author[0000-0001-7594-8072]{Yongseok Lee}
\affil{Korea Astronomy and Space Science Institute, Daejeon 34055, Republic of Korea}
\affil{School of Space Research, Kyung Hee University, Yongin 17104, Republic of Korea}

\author[0009-0000-5737-0908]{Dong-Joo Lee}
\affil{Korea Astronomy and Space Science Institute, Daejeon 34055, Republic of Korea}
\affil{Department of Astronomy and Space Science, Chungbuk National University, Cheongju 28644, Republic of Korea}

\author[0000-0002-6982-7722]{Byeong-Gon Park}
\affil{Korea Astronomy and Space Science Institute, Daejeon 34055, Republic of Korea}

\begin{abstract}
Blue large-amplitude pulsators (BLAPs), a recently classified type of variable stars, are evolved objects likely formed through interactions between stars in a binary system. However, only two BLAPs with stellar companions have been discovered to date. This paper presents photometric data from the Korea Microlensing Telescope Network (KMTNet) for three BLAPs located in the direction of the Galactic bulge: OGLE-BLAP-006, OGLE-BLAP-007, and OGLE-BLAP-009. The data were collected over eight consecutive years, beginning in 2016, with a high cadence of approximately 15 minutes. Frequency analysis of light variations revealed OGLE-BLAP-006 as a multimode pulsator with a dominant frequency of 37.88 day$^{-1}$ and two new frequencies of 38.25 and 35.05 day$^{-1}$. In contrast, OGLE-BLAP-007 and OGLE-BLAP-009 exhibit single-mode pulsation. By combining the KMTNet data with archival OGLE observations, we investigated pulsation timing variations of the BLAPs using an $O-C$ diagram to identify the light travel time effect caused by the orbital motion of their companions. We found that OGLE-BLAP-006, with no evidence of close companions, has two wide-orbit companions with orbital periods of approximately 4,700 and 6,300 days, making it the third known BLAP in a stellar system; however, no companions were found for OGLE-BLAP-007 and OGLE-BLAP-009. Furthermore, we identified seven other BLAP candidates with wide companions using OGLE data, suggesting that such systems are relatively common. We propose that a BLAP with a wide companion may be a merger remnant of an inner close binary within a hierarchical triple system.
\end{abstract}

\section{Introduction \label{sec_intro}}
Blue large-amplitude pulsators (BLAPs) are a recent class of variable stars introduced by \citet{pietrukowicz2017}, who reported the first 14 BLAPs using data from the Optical Gravitational Lensing Experiment \citep[OGLE;][]{udalski2015}. These stars are characterized by their sawtooth-shaped light curves with amplitudes over 0.2 mag, similar to Cepheids and RR Lyr stars. However, BLAPs have unusually short pulsation periods of 3--75 minutes and are hot objects discovered in a rarely populated area between subdwarf B-type stars and B-type main-sequence stars on the Hertzsprung-Russell (H-R) diagram, with effective temperatures of 25,000--34,000 K and surface gravities of $\log\,g$ = 4.2--5.7 \citep{pietrukowicz2024}. As expected theoretically, their measured gravities correlate linearly with pulsation periods on logarithmic scales; higher gravity results in shorter periods. As deduced from the light curves, BLAPs are generally assumed to be single-mode pulsators excited in the fundamental radial mode. However, some stars exhibit different characteristics. For example, the pulsation period of OGLE-BLAP-009 corresponds to a first-overtone radial mode \citep{bradshaw2024}, while the four BLAPs, namely SMSS-BLAP-1, OGLE-BLAP-001, OGLE-BLAP-030, and ZGP-BLAP-08, demonstrate multiple modes \citep{chang2024, pietrukowicz2024, pigulski2024}.

Following their initial discovery in 2017, the number of BLAPs has increased significantly to over 100 owing to large-scale photometric surveys. In particular, the OGLE group discovered 88 BLAPs: the prototype object OGLE-BLAP-001 in the Galactic disk, 13 BLAPs in the inner Galactic bulge \citep{pietrukowicz2017}, 20 new BLAPs in the Galactic disk \citep{borowicz2023a}, 31 BLAPs in the outer Galactic bulge \citep{borowicz2023b}, and an additional 23 BLAPs in the inner Galactic bulge \citep{pietrukowicz2024}. In addition, \citet{kupfer2019} discovered four BLAPs with high gravities of $\log\,g$ = 5.3--5.7 using high-cadence observations with the Zwicky Transient Facility \citep[ZTF;][]{bellm2019}, while \citet{mcwhirter2022} reported 22 candidates by inspecting the ZTF light curves, four of which were subsequently confirmed as BLAPs by \citet{borowicz2023a}. Additional discoveries include three BLAPs from the OmegaWhite Survey \citep{ramsay2022}, TMTS-BLAP-1 from the Tsinghua University-Ma Huateng Telescopes for Survey \citep{lin2023}, and SMSS-BLAP-1 from the SkyMapper Southern Survey \citep{chang2024}.

BLAPs are considered as core remnants of red giants whose envelopes were stripped away or suffered significant mass loss, similar to subdwarf B-type stars located nearby in the H-R diagram \citep{heber2016}. Such mass loss is unlikely to occur during the evolution of a single, isolated, low-mass star within the Hubble time \citep{pietrukowicz2017, corsico2019}. Therefore, BLAPs are believed to be formed through interactions between stars in a binary system, such as Roche lobe overflow or common envelope ejection \citep{byrne2021}.  Several theoretical models have been proposed and are still under debate. These models can be categorized into three types based on stellar structure: shell H-burning pre-white dwarfs with low-mass He cores of $\sim$0.3$M_\sun$ \citep{pietrukowicz2017,byrne2018,romero2018,kupfer2019,byrne2020}, core He-burning subdwarfs \citep{wu2018}, and shell He-burning subdwarfs \citep{xiong2022,lin2023}. Other formation scenarios suggest that BLAPs could be surviving companions with a He-burning core from single-degenerate Type Ia supernovae \citep{meng2020} or remnants from the merger of a He-core white dwarf and a low-mass main-sequence star \citep{zhang2023}. These two models are also related to the evolution of binary systems, but they predict BLAPs as single stars without companions. Recently, \citet{kolaczek2024} proposed a new model involving the merger of extremely low-mass (ELM) double white dwarfs, potentially leading to the formation of magnetic BLAPs.

Binary interaction models anticipate that many BLAPs will exist in binary systems, with short orbital periods of $\sim$40 days for low-mass pre-white dwarf models \citep{byrne2021} and long periods of $\sim$1,400 days for shell He-burning subdwarf models \citep{xiong2022}. By contrast, some models suggest the existence of single BLAPs. Thus, the binary fraction and orbital period distribution can provide valuable constraints on BLAP formation channels. However, a systematic search for binarity is yet to be performed, and most BLAPs remain unknown. Two BLAPs have been found in binaries: HD 133729 with an orbital period of $\sim$23 days \citep{pigulski2022} and TMTS-BLAP-1 with a period of $\sim$1,580 days \citep{lin2023}. These binaries were discovered through the pulsation timing variation method, which detects the light-travel time (LTT) effect caused by the orbital motion of companion stars in the $O-C$ diagram; $O-C$ refers to the difference between the observed and calculated times of maximum light. Eclipses have not yet been identified in the light curves of BLAPs, implying a low fraction of short-period close binaries.

We initiated a systematic survey to examine the binary fraction of BLAPs, using data from the Korea Microlensing Telescope Network \citep[KMTNet;][]{kim2016a}. KMTNet is a photometric system comprising three 1.6\,m telescopes, each equipped with an 18k $\times$ 18k pixel mosaic CCD camera, offering a wide field of view of 2.0 $\times$ 2.0 square degrees. These telescopes operate at three observatories in the Southern Hemisphere, enabling continuous 24-hour monitoring of the southern sky: the Cerro Tololo Inter-American Observatory (CTIO) in Chile, the South African Astronomical Observatory (SAAO) in South Africa, and the Siding Spring Observatory (SSO) in Australia. Since 2016, KMTNet has been monitoring the Galactic bulge with high cadence from late February to late October annually to search for exoplanets through gravitational microlensing. These data are also well suited for identifying BLAPs in binary systems across a range of orbital periods from short to long by detecting eclipses and pulsation timing variations. Many of the 36 BLAPs discovered in the inner Galactic bulge by \citet{pietrukowicz2017, pietrukowicz2024} are located in the 27 fields observed by KMTNet \citep{kim2018}, serving as key targets for our investigation.

As the first work of our survey, this paper presents photometric results for three BLAPs: OGLE-BLAP-006, OGLE-BLAP-007, and OGLE-BLAP-009. We report that OGLE-BLAP-006 is the third BLAP with stellar companions, following previously identified BLAPs, namely HD 133729 and TMTS-BLAP-1.

\section{Observation and Data Reduction \label{sec_obsdata}}
We used KMTNet data collected over eight consecutive years, starting in 2016. Data for 2020 decreased considerably due to the shutdown of two host sites, CTIO and SAAO, during the COVID-19 pandemic. Of the 27 KMTNet Bulge fields \citep{kim2018}, we observed OGLE-BLAP-006 and OGLE-BLAP-007 in fields BLG02 and BLG42, whereas OGLE-BLAP-009 was observed in fields BLG03 and BLG43. The prime fields, BLG02 and BLG03, are slightly shifted and nearly overlap with BLAP42 and BLAP43, respectively. These four fields were observed twice per hour, mainly in the $I$ band with a 60-second exposure and occasionally in the $V$ band with a 90-second exposure. The combined data from the two overlapping fields yield a high cadence of 15 minutes.

Difference image analysis was applied to the KMTNet data using the Python package {\tt pyDIA} \citep{albrow2017}. We converted output fluxes from {\tt pyDIA} into standard magnitudes using the empirical relations for each passband (E.-C. Sung, priv. comm.) derived from observations of photometric standard stars \citep{landolt1992} at the KMTNet-CTIO site. The conversion error was approximately 0.05 mag. The output times from {\tt pyDIA}, expressed in heliocentric Julian date, were converted to barycentric Julian date (BJD$_{\rm TDB}$) using the Python code {\tt convert\_times.py} (\url{https://github.com/WarwickAstro/time-conversions}). In addition, we corrected for a few seconds of time delay from a large-format mechanical shutter \citep{kim2016b}, which varies across the telescope's focal plane. Poor-quality data points, such as those with chi-square values greater than 2.0, full width at half maximum exceeding 5.0 arcsec, and background fluxes above 10,000 ADU, were excluded from the analysis. We removed outliers that deviated significantly from periodic light variations due to pulsation, as they appeared randomly and did not result from the binary star eclipses.

In addition, we used OGLE $I$-band data to examine long-term timing variations of the three BLAPs. The downloaded time-series photometric data from the OGLE collection of variable stars\footnote{\url{https://www.astrouw.edu.pl/ogle/ogle4/OCVS/BLAP}} are formatted as BJD$_{\rm TDB} - 2,450,000$, standard magnitudes, and magnitude errors. The data cover three phases for OGLE-BLAP-006, from OGLE-II to OGLE-IV, spanning 23 years from 1997 to 2019. In contrast, data for OGLE-BLAP-007 and OGLE-BLAP-009 were collected from OGLE-III and OGLE-IV, spanning 19 years from 2001 to 2019. Similar to the KMTNet data, outliers in the OGLE data were excluded from the analysis.

\section{OGLE-BLAP-006 \label{sec_blap06}}
\subsection{Light Curve and Frequency Analysis \label{subsec_blap06_LC}}
\begin{figure*}
\center\includegraphics[scale=0.45]{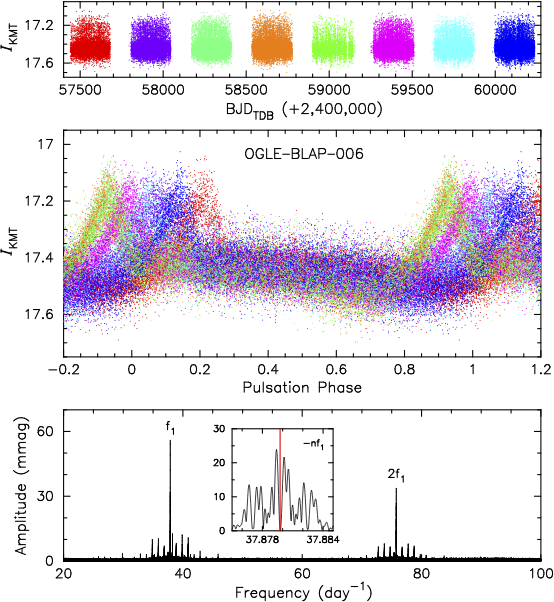} \hskip 3mm \includegraphics[scale=0.44]{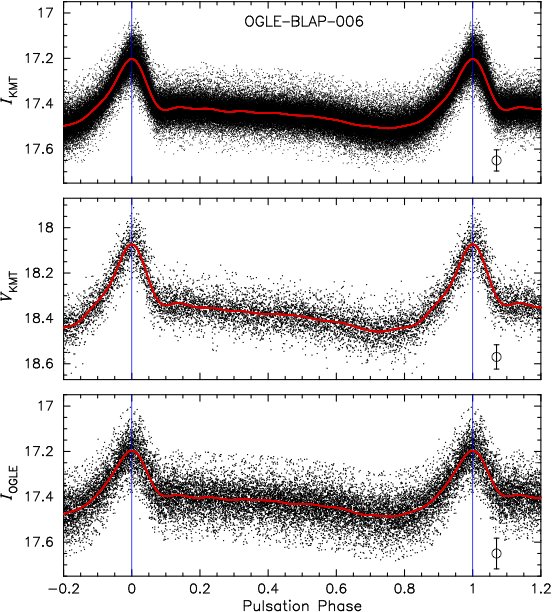}
\caption{Light variations of OGLE-BLAP-006. (Left) Top panel displays the KMTNet $I$-band data, spanning eight years from 2016, with distinct colors for each year. The phase diagram in the middle reveals notable annual changes in the phase of maximum brightness, indicating pulsation timing variations. The pulsation phases were computed using the primary frequency $f_1$ from the Fourier amplitude spectrum in the bottom panel. The inset offers a close-up view of prewhitening results around the frequency $f_1$, marked by a vertical red line, highlighting prominent side peaks due to the timing variations. $-nf_1$ represents the prewhitening process that subtracts light variations linked to $f_1$ and its harmonics. (Right) Phase diagrams corrected for timing variations using the $O-C$ values from Figure \ref{fig_blap06_oc} are shown for three data sets: from the top, the KMTNet $I$, $V$, and OGLE $I$ bands. Red lines represent synthetic fitting curves, with fitting error bars in the bottom right of each panel.\label{fig_blap06_lc}}
\end{figure*}

The light variations of OGLE-BLAP-006 are displayed in Figure \ref{fig_blap06_lc}. We examined the periodicity of these variations through a multiple-frequency analysis \citep{kim2010} utilizing the discrete Fourier transform and least-squares fitting, as follows:
\begin{equation}
m = m_0 + \sum_{i=1}^k A_i \sin(2\pi f_i t + \phi_i)
\end{equation}
\noindent
where $m$ and $t$ represent the observed magnitude and time, respectively. The amplitude $A_i$ and phase $\phi_i$ for each frequency $f_i$ and the mean magnitude $m_0$ were obtained through nonlinear fitting using the Levenberg-Marquardt method \citep{press1986}.

The phase diagram of OGLE-BLAP-006, in the middle-left panel of Figure \ref{fig_blap06_lc}, appears as a poorly defined and dispersed light curve resembling the uncorrected TMTS-BLAP-1 curve presented in Extended Data Figure 5 by \citet{lin2023}. The phase of maximum light varies significantly each year, shifting from approximately +0.21 in 2016 (red dots; BJD$_{\rm TDB} \approx$ 2,457,500) to approximately $-$0.08 in 2019 (orange dots; BJD$_{\rm TDB} \approx$ 2,458,600) and then back to approximately +0.13 in 2023 (blue dots; BJD$_{\rm TDB} \approx$ 2,460,100), while the maximum brightness itself remains nearly constant. These features indicate variations in timing. The notable side peaks around the primary frequency in the Fourier amplitude spectrum, shown in the inset of the bottom-left panel, are due to these timing variations. Consequently, we corrected these variations using the $O-C$ values discussed in the following subsection. The right panels illustrate the resulting coherent light curves. The observed sawtooth-shaped light curves fit reasonably well with the synthetic curves employing harmonics up to the 10th order.

\begin{figure*}
\center\includegraphics[scale=0.46]{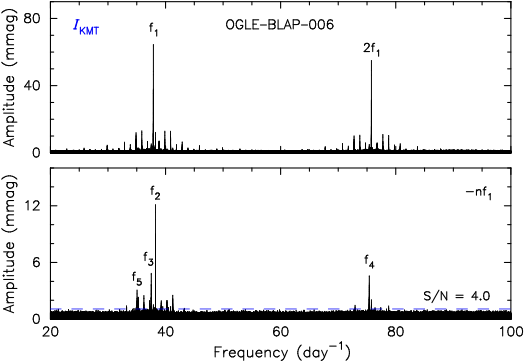}\hskip 3mm\includegraphics[scale=0.46]{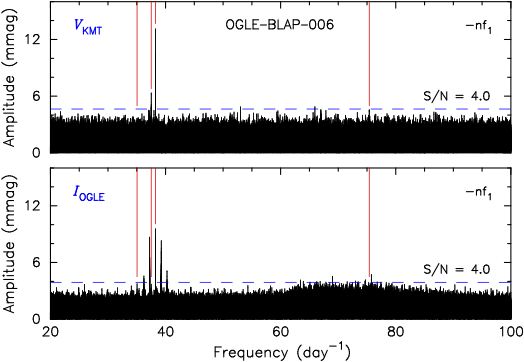}
\caption{Fourier amplitude spectra of OGLE-BLAP-006 after correcting for timing variations. (Left) Upper panel displays the KMTNet $I$-band spectrum, showing higher amplitudes than those obtained before the timing correction in Figure \ref{fig_blap06_lc}. The prewhitening process in the lower panel reveals four new frequencies, $f_2$ to $f_5$. Frequencies with amplitudes above the horizontal blue dashed line of S/N = 4.0 are 1.0 day$^{-1}$ aliasing sidelobes and close frequencies. (Right) Upper and lower panels show prewhitening results for the KMTNet $V$ and OGLE $I$ bands, respectively. The vertical red lines indicate frequencies $f_2$ to $f_5$ identified in the KMTNet $I$-band data. The OGLE spectrum exhibits 1.0 day$^{-1}$ aliasing sidelobes with significantly higher amplitudes than those in the KMTNet.\label{fig_blap06_ps}}
\end{figure*}

Figure \ref{fig_blap06_ps} presents the Fourier amplitude spectra after correcting for timing variations. The upper-left panel displays the KMTNet $I$-band spectrum, where the primary frequency $f_1$ and its harmonic $2f_1$ demonstrate comparable amplitudes with a ratio of approximately 0.83. The lower-left panel shows the prewhitening results after subtracting light variations associated with $f_1$ and its harmonic frequencies up to the 10th order. We detected more than 20 new frequencies with an amplitude signal-to-noise ratio (S/N) exceeding the empirical criterion of 4.0 \citep{breger1993} in the 20--100 day$^{-1}$ range using this iterative process. Table \ref{tab_blap06_freq} lists 11 frequencies with amplitudes above 0.0030 mag considered in this study, all having an S/N above 10.0. Frequencies with lower amplitudes but an S/N above 4.0 were identified as close frequencies. We adopted a frequency resolution criterion of 1.0/$\Delta$T = 0.00036 day$^{-1}$, where $\Delta$T (approximately 2,800 days) represents the total observation period. The right panels of Figure \ref{fig_blap06_ps} indicate that the new frequency $f_2$ detected in the KMTNet $I$ band also appears in KMTNet $V$ and OGLE $I$ bands. The Fourier spectrum for OGLE data was significantly affected by the 1.0 day$^{-1}$ aliasing effect from single-site observations.

\begin{deluxetable}{lcl}
\tablewidth{0pt} \tablecolumns{3}
\tablecaption{Results of the multiple frequency analysis for OGLE-BLAP-006 in the 20--100 day$^{-1}$ range \label{tab_blap06_freq}}
\tablehead{ \colhead{Frequency (day$^{-1}$)} & \colhead{Amplitude (mag)} & \colhead{Remark} }
\startdata
\,\,\,\,$f_1$ = 37.8797546(6) & 0.0663(2) & Independent \\
\,2$f_1$ = 75.7595096(8) & 0.0553(2) & Harmonic  \\ 
\,\,\,\,$f_2$ = 38.2451416(47) & 0.0118(2) & Independent \\
$f_{2,1}$ = 38.2447288(82) & 0.0066(2) & Close to $f_2$ \\
$f_{2,2}$ = 38.2466622(72) & 0.0060(2) & Close to $f_2$ \\
\,\,\,\,$f_3$ = 37.5143442(95) & 0.0052(2) & 2$f_1 - f_2$ \\
\,\,\,\,$f_4$ = 75.3941022(104) & 0.0050(2) & 3$f_1 - f_2$ \\
$f_{3,1}$   = 37.5148065(148) & 0.0033(2) & 2$f_1 - f_{2,1}$ \\
\,\,\,\,$f_5$ = 35.0472846(116) & 0.0030(2) & Independent \\
$f_{2,3}$ = 38.2485397(148) & 0.0033(2) & Close to $f_2$ \\
$f_{4,1}$ = 75.3945236(168) & 0.0031(2) & 3$f_1 - f_{2,1}$ \\
\enddata
\tablenotetext{}{{\bf Note.} Values in parentheses represent errors in the last digits. If we change the frequency resolution criterion to 1.5/$\Delta$T = 0.00054 day$^{-1}$ \citep{loumos1978}, frequencies $f_{2,1}$, $f_{3,1}$, and $f_{4,1}$ would be classified as unresolved.}
\end{deluxetable}

\begin{figure}
\center\includegraphics[scale=0.44]{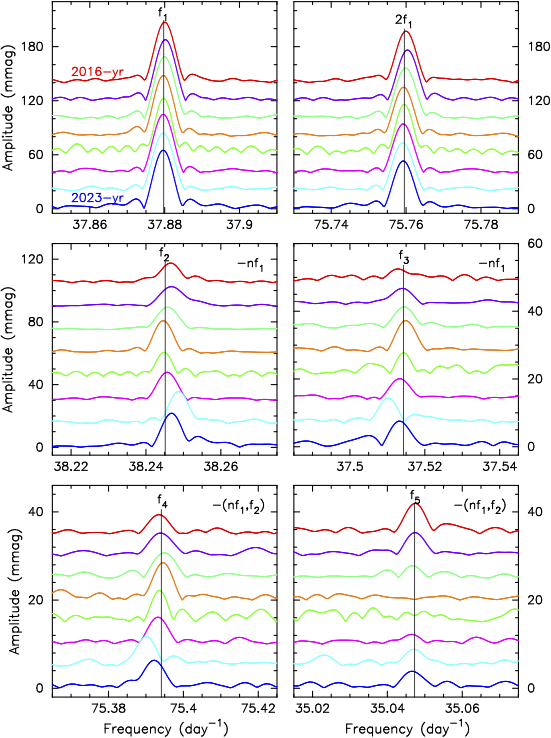}
\caption{Fourier spectra of OGLE-BLAP-006 from KMTNet $I$-band data separated by year, with an arbitrary offset on the y-axis. These zoomed-in views focus on six frequencies: $f_1$, 2$f_1$, and $f_2$ to $f_5$. Notable frequency changes appear in $f_2$, $f_3$, and $f_4$. \label{fig_blap06_ps_yr}}
\end{figure}

These results reveal two notable features. The first is the presence of several frequencies close to $f_2$ (denoted as $f_{2,j}$; see Table  \ref{tab_blap06_freq}) and their combination frequencies such as $f_{3,1}$ = $2f_1 - f_{2,1}$. Close frequencies near $f_2$ were also detected in the OGLE $I$-band data: 38.2487 day$^{-1}$ with an amplitude of 0.0115 mag, 38.2474 day$^{-1}$ with an amplitude of 0.0078 mag, and 38.2497 day$^{-1}$ with an amplitude of 0.0066 mag. To investigate the origin of these close frequencies, we conducted a frequency analysis of the KMTNet $I$-band data by year. The resulting spectra (Figure \ref{fig_blap06_ps_yr}) show that $f_2$ shifts considerably from 38.2446 day$^{-1}$ in 2019 to 38.2487 day$^{-1}$ in 2022, whereas $f_1$, 2$f_1$, and $f_5$ remain nearly constant. In contrast, $f_3$ and $f_4$ vary inversely to $f_2$, resembling a mirror image. These frequency shifts near $f_2$ suggest that OGLE-BLAP-006 may have experienced mode-switching between a few closely spaced frequencies with different pulsation modes.

The second key observation is the exact equidistance of $f_2$ and $f_3$ from the dominant frequency $f_1$; that is, $f_2 - f_1$ = $f_1 - f_3$ = 0.36540 $\pm$ 0.00001 day$^{-1}$, with a significantly small deviation of $ | \, (f_2 - f_1) - (f_1 - f_3) \, |$ = 0.00002 $\pm$ 0.00001 day$^{-1}$. Two possible explanations for this triplet include sinusoidal amplitude variability and the combination-mode hypothesis \citep{breger2006}. Equidistant frequency triplets resulting from sinusoidal amplitude variability have been identified in rapidly oscillating Ap stars and, more recently, in two BLAPs, OGLE-BLAP-001 and ZGP-BLAP-08 \citep{pigulski2024}. While this variability predicts two side peaks of equal amplitude, the $f_3$ of OGLE-BLAP-006 exhibits a significantly smaller amplitude, less than half that of $f_2$. Therefore, the frequency triplet in OGLE-BLAP-006 is unlikely to be attributed to sinusoidal amplitude variability. Instead, we suggest $f_3$ as a combination frequency, given by $f_3$ = $2f_1 - f_2$, as shown in Table \ref{tab_blap06_freq}. The existence of a similar combination frequency $f_4$ = $3f_1 - f_2$ supports our conclusion. Consequently, OGLE-BLAP-006 is a multimode pulsator with two new independent frequencies, $f_2$ = 38.2451 day$^{-1}$ and $f_5$ = 35.0473 day$^{-1}$, alongside the dominant frequency $f_1$ = 37.8798 day$^{-1}$.

\subsection{Pulsation Timing Variations \label{subsec_blap06_OC}}
The pulsating star's distance from us in a binary system changes due to its orbital motion, assuming that the orbital inclination is not zero. This results in the observed times of maximum light being periodically delayed or advanced. This LTT effect is widely used to search for stellar companions \citep[for example,][]{murphy2014,prudil2019,otani2022,vaulato2022,dholakia2025}, known as the pulsation timing variation or phase modulation method.

We examined the pulsation timing variations of OGLE-BLAP-006 using the $O-C$ diagram. The $I$-band time-series data were divided into several short segments, and least-squares fitting was performed for each segment using Equation (1). The segment size was carefully selected because a shorter segment would provide insufficient data for fitting, whereas overly long segments could hinder the detection of short-term LTT variations. We found a segment length of one year to be reasonable for the OGLE data. Accordingly, a 0.5-year segment was selected for KMTNet despite having more high-cadence data. Segments with insufficient data points or high fitting errors were excluded from the subsequent analyses.

Phase $\phi_i$, one of the fitting parameters, was determined for 11 fixed frequencies, $f_1$ through $10f_1$ and $f_2$; including $f_2$ was found to have little impact on our results. $\phi_1$ for the primary frequency $f_1$, corresponding to a representative pulsation period, was then converted to the times of maximum light $T_{\rm max}$. We adopted $T_{\rm max}$ closest to the midpoint of each segment as the observed time of maximum light, i.e., $O$. The calculated times of maximum light, $C$, were derived from the following linear equation:
\begin{equation}
C = T_0 + P \, E
\end{equation}
\noindent
where the pulsation period $P$ (or the reciprocal of $f_1$) is assumed constant. The reference time $T_0$ was arbitrarily set to BJD$_{\rm TDB}$ = 2,457,513.87134, one observed time of maximum light from the KMTNet $I$-band data. Epoch $E$ denotes the integer value of $(O - T_0)/P$, representing the number of pulsation cycles since $T_0$. We adjusted the epochs of the first four $O-C$ data points, which exceeded a single pulsation cycle, to ensure continuity in the variation trend. The observed times of maximum light, epochs, and $O-C$ values are presented in the Appendix Table \ref{tab_timing}.

\begin{figure}
\center\includegraphics[scale=0.44]{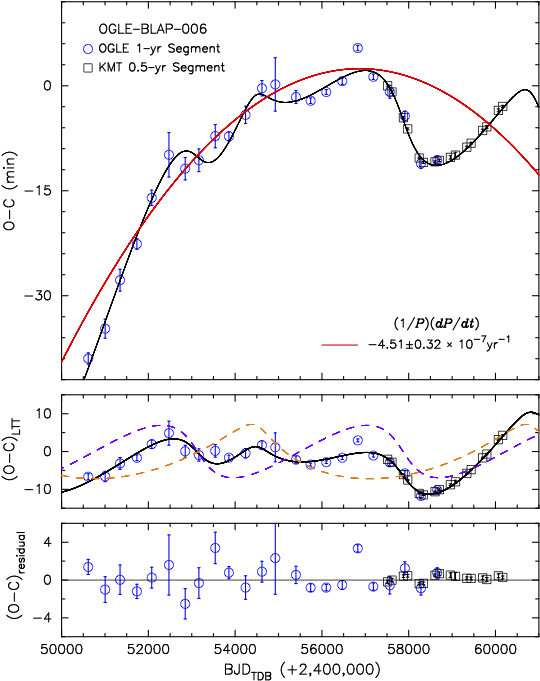}
\caption{$O-C$ diagram for the pulsation period of OGLE-BLAP-006. The OGLE and KMTNet data were divided into 1.0-year and 0.5-year segments, respectively, to derive the observed times of maximum light. The red curve in the top panel represents a model parabola with a period change rate of ($1/P$)($dP/dt$) = $-$4.51 $\times$ 10$^{-7}$ year$^{-1}$. The black curve is of the model that combines the parabola with two LTT terms; the residuals of the model fit are shown in the bottom panel. The middle panel displays the two LTT terms: a blue dashed curve for the inner companion, a red dashed curve for the outer companion, and a thick black curve for their combined model. The error bars in the $O-C$ values, derived from phase errors ($\sigma \phi_1$), were doubled, as recommended by \citet{handler2000}. \label{fig_blap06_oc}}
\end{figure}

As shown in the top panel of Figure \ref{fig_blap06_oc}, the resulting $O-C$ variations of OGLE-BLAP-006 cannot be interpreted as a single parabola indicative of a linear period change, contrasting sharply with those of OGLE-BLAP-007 and OGLE-BLAP-009 (see Figure \ref{fig_blap0709_oc}). After several trials, we reasonably modeled the timing variations by combining a quadratic polynomial with two LTT terms as follows: 
\begin{equation}
(O-C)_{\rm model} = \Delta T_0 + \Delta P \, E + A \, E^2 + \tau_{\rm in} + \tau_{\rm out}
\end{equation}
\noindent
where $\Delta T_0$ and $\Delta P$ denote offsets from the initial values of $T_0$ and $P$, respectively. The quadratic coefficient $A = 0.5 P (dP/dt)$ indicates the rate of period change. $\tau_{\rm in}$ and $\tau_{\rm out}$ represent the LTTs of OGLE-BLAP-006 due to the orbital motions of the inner and outer companions, respectively. As formulated by \citet{irwin1952}, each LTT term includes five orbital parameters: orbital period $P_{\rm orb}$, projected semi-major axis of the pulsator $a_{\rm pul} \sin i$, eccentricity $e$, argument of periastron $\omega$, and time of periastron passage $T_{\rm peri}$. Thus, we must determine 13 unknown parameters through modeling.

Model fitting was initially performed using the Python function ${\tt differential\_evolution}$ from the ${\tt scipy.optimize}$ module \citep{storn1997} to determine the global minimum. Based on these optimization results, we set the priors and explored the parameter space using the Python package {\tt emcee} \citep{foremanmackey2013}, an ensemble sampler for Markov chain Monte Carlo (MCMC) methods, to estimate parameter uncertainties. We adopted the median of the marginal posterior distributions as the best-fit estimate for each parameter and calculated the uncertainty as the average of the 16th and 84th percentiles.

\begin{deluxetable}{lcc}
\tablewidth{0pt} \tablecolumns{3}
\tablecaption{MCMC estimates for the two LTT terms of OGLE-BLAP-006 \label{tab_blap06_ltt}}
\tablehead{ \colhead{Parameters} & \colhead{$\tau_{\rm in}$} & \colhead{$\tau_{\rm out}$} }
\startdata
$P_{\rm orb}$ (day)                         & 4,737 $\pm$ 137 & 6,320 $\pm$ 301\\
$a_{\rm pul} \sin i$ (AU)                  & 0.95 $\pm$ 0.05 & 0.90 $\pm$ 0.09 \\ 
$e$                                                 & 0.48 $\pm$ 0.05 & 0.63 $\pm$ 0.07 \\ 
$\omega$ (degree)                         & 180 $\pm$ 7 & 116 $\pm$ 13 \\                          
$T_{\rm peri}$ (BJD$_{\rm TDB}$) & 2,457,873 $\pm$ 66 & 2,454,524 $\pm$ 178 \\
\enddata
\end{deluxetable}

\begin{figure}
\center\includegraphics[scale=0.44]{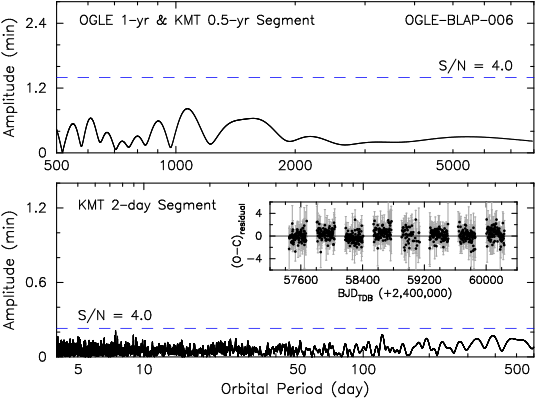}
\caption{Fourier spectra of the $O-C$ residuals for OGLE-BLAP-006. The upper panel shows the spectrum for long orbital periods, from 500 to 8,000 days, derived from the residuals for OGLE's 1.0-year segments and KMTNet's 0.5-year segments. The lower panel displays the spectrum for short-period regions, from 4 to 600 days, obtained using the $O-C$ residuals (shown in the inset) for KMTNet's 2.0-day segments. \label{fig_blap06_ocres}}
\end{figure}

The quadratic coefficient $A$ was estimated to be $-4.30 \pm 0.30 \times 10^{-13}$ days, corresponding to a period change rate of $(1/P)(dP/dt) = -4.51 \pm 0.32 \times 10^{-7}$ year$^{-1}$. This secular change in the pulsation period is probably due to stellar evolution. Our estimate, based on timing variation analysis, closely matches the previous value of $-3.32 \pm 0.37 \times 10^{-7}$ year$^{-1}$ reported by \citet{pietrukowicz2024}, derived through a period-change analysis using only OGLE data. The MCMC estimates for the LTT terms are listed in Table \ref{tab_blap06_ltt}. The predicted LTT variations appear in the middle panel of Figure \ref{fig_blap06_oc}, indicated by the blue and red dashed curves for $\tau_{\rm in}$ and $\tau_{\rm out}$, respectively. The combined model predictions, represented by the black lines in the top and middle panels, closely match the observed data. No significant features are observed in the residuals (bottom panel of Figure \ref{fig_blap06_oc}) and their Fourier spectrum (upper panel of Figure \ref{fig_blap06_ocres}). However, our estimates exhibit some uncertainty due to inadequate observation of long-term timing variations, necessitating additional data over several years to refine parameter values.

KMTNet $I$-band data were continuously collected from three host sites with a high cadence of 15 minutes, allowing for detecting LTT variations caused by stellar companions with orbital periods as short as a few days. To identify these short-period companions, we divided the KMTNet data into 2.0-day segments, the minimum interval providing adequate data for fitting. The lower panel of Figure \ref{fig_blap06_ocres} displays the Fourier amplitude spectrum of the $O-C$ residuals for these segments. The inset in the lower panel shows the residuals obtained by subtracting the long-term timing variations predicted by the model in Equation (3). In the Fourier spectrum for orbital periods between 4 and 600 days, we found no significant peaks with LTT amplitudes exceeding 0.23 minutes (or 14 seconds), applying a detection criterion of S/N > 4.0.

Consequently, we propose OGLE-BLAP-006 as the third BLAP known to have stellar companions, following HD 133729 \citep{pigulski2022} and TMTS-BLAP-1 \citep{lin2023}. Our discovery of companion stars with orbital periods spanning several thousand days in OGLE-BLAP-006 is surprising and theoretically unexpected. We discuss these wide-orbit companions around BLAPs in Section \ref{sec_discussion}.

\section{OGLE-BLAP-007 and OGLE-BLAP-009 \label{sec_blap0709}}
\subsection{Light Curve and Frequency Analysis \label{subsec_blap0709_LC}}
\begin{figure*}
\center\includegraphics[scale=0.437]{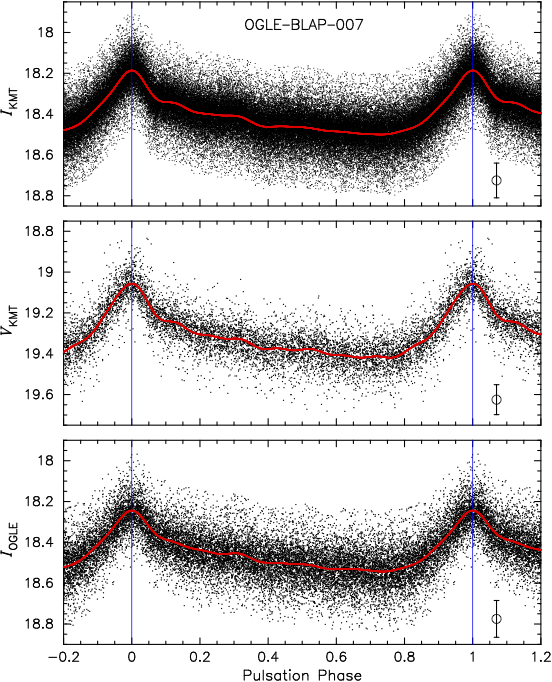} \hskip 3mm \includegraphics[scale=0.437]{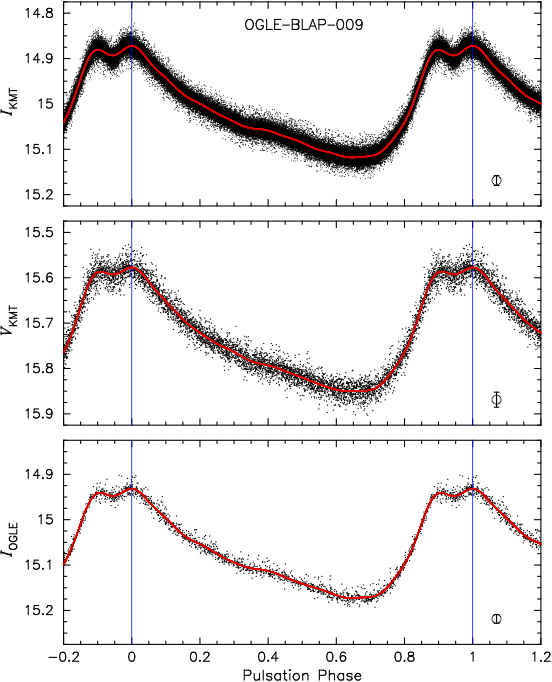}
\caption{Phase diagrams of OGLE-BLAP-007 (left) and OGLE-BLAP-009 (right), corrected for timing variations using the $O-C$ values from Figure \ref{fig_blap0709_oc}. The pulsation phases are calculated using the primary frequency $f_1$ obtained from the Fourier spectra in Figure \ref{fig_blap0709_ps}. The other descriptions are the same as those for the right panels of Figure \ref{fig_blap06_lc}. \label{fig_blap0709_lc}}
\end{figure*}

The light curves for OGLE-BLAP-007 and OGLE-BLAP-009 are shown in Figure \ref{fig_blap0709_lc}. These coherent phase diagrams are corrected for timing variations using the $O-C$ diagram in Figure \ref{fig_blap0709_oc}. Previous studies \citep{macfarlane2017, mcwhirter2020, ramsay2022, bradshaw2024} found that OGLE-BLAP-009 exhibits double peaks with a small dip at maximum brightness. This feature is captured well by a harmonic fit up to the 10th order, illustrated by the red lines in Figure \ref{fig_blap0709_lc}. A similar feature was identified by \citet{pigulski2022} in the 20-second cadence TESS data for HD 133729, where a brief tiny dip occurs shortly after maximum light. These unusual dips may result from a shock at the minimum radius or resonance between the fundamental and second overtone radial modes \citep{bradshaw2024}.

\begin{figure*}
\center\includegraphics[scale=0.45]{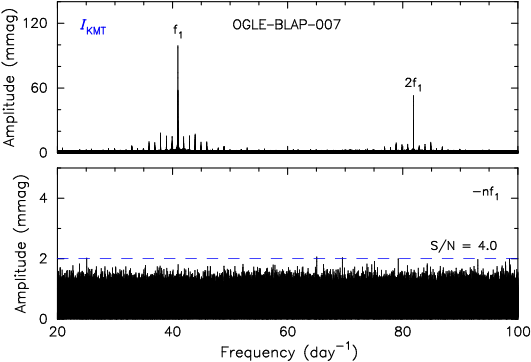} \hskip 3mm \includegraphics[scale=0.45]{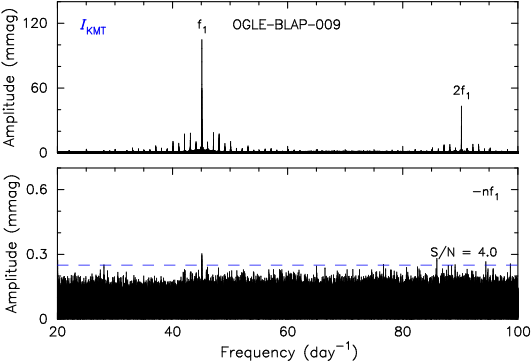}
\caption{Fourier amplitude spectra of OGLE-BLAP-007 (left) and OGLE-BLAP-009 (right) for the KMTNet $I$-band data, corrected for timing variations. The lower panels display the prewhitening results. A peak near $f_1$ in the lower-right panel is identified as an unresolved frequency. \label{fig_blap0709_ps}}
\end{figure*}

We performed a Fourier analysis on the KMTNet $I$-band data. Similar to the results for OGLE-BLAP-006 in the bottom-left panel of Figure \ref{fig_blap06_lc}, the prewhitening process for OGLE-BLAP-007 and OGLE-BLAP-009 revealed side peaks around the primary frequency, although at significantly lower amplitudes. These side peaks were attributed to timing variations, prompting a reanalysis using data corrected for these variations. Figure \ref{fig_blap0709_ps} presents the amplitude spectra from this reanalysis. The primary frequencies $f_1$ = 40.9306 day$^{-1}$ for OGLE-BLAP-007 and $f_1$ = 45.0912 day$^{-1}$ for OGLE-BLAP-009, along with their respective harmonics ($2f_1$), are clearly identified in the upper panels. The lower panels show no additional significant peaks after prewhitening, confirming that both stars are single-mode pulsators \citep{mcwhirter2020}. The significance of several peaks with amplitudes just above the detection threshold (S/N = 4.0) remains questionable, as extensive time-series data from long-term surveys such as KMTNet can increase the likelihood of spurious detections \citep{baran2015}.

\subsection{Pulsation Timing Variations \label{subsec_blap0709_OC}}
\begin{figure*}
\center\includegraphics[scale=0.45]{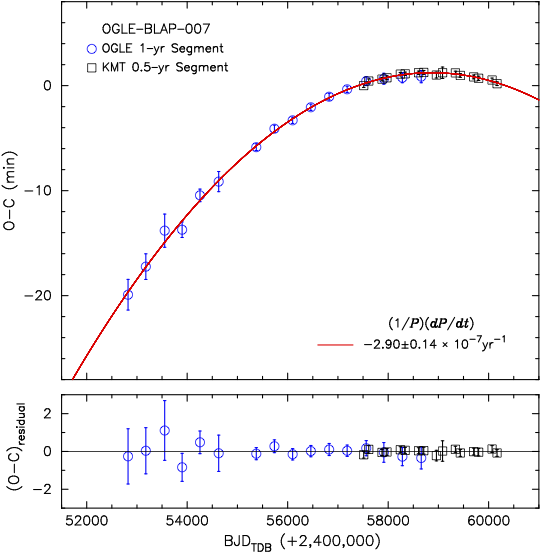} \hskip 3mm \includegraphics[scale=0.45]{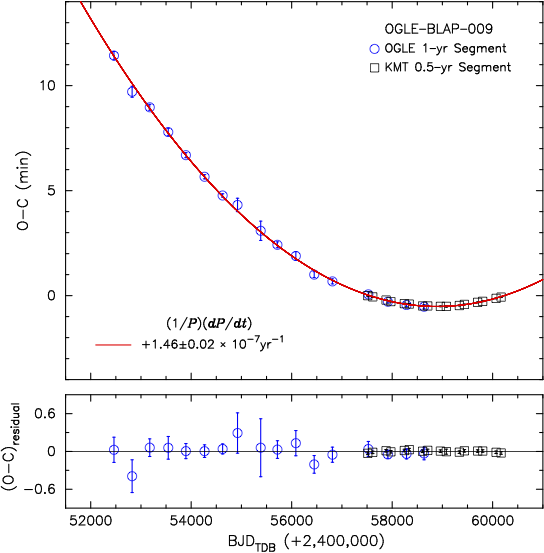}
\caption{$O-C$ diagram for the pulsation periods of OGLE-BLAP-007 (left) and OGLE-BLAP-009 (right). Red lines in the upper panels represent the parabolic fits; the corresponding residuals are shown in the lower panels. \label{fig_blap0709_oc}}
\end{figure*}

We examined the pulsation timing variations for OGLE-BLAP-007 and OGLE-BLAP-009, similarly to OGLE-BLAP-006. The resulting $O-C$ diagrams, using timing values from Appendix Table \ref{tab_timing}, appear in Figure \ref{fig_blap0709_oc}. The timing variations of both stars fit a parabolic curve well, indicating a linear change in the pulsation period over time. We estimated the period change rates $(1/P)(dP/dt)$ as $-2.90 \pm 0.14 \times 10^{-7}$ year$^{-1}$ for OGLE-BLAP-007 and $+1.46 \pm 0.02 \times 10^{-7}$ year$^{-1}$ for OGLE-BLAP-009. These estimates align with earlier results reported by \citet{pietrukowicz2024}: $-3.41 \pm 0.45 \times 10^{-7}$ year$^{-1}$ for OGLE-BLAP-007 and $+1.50 \pm 0.05 \times 10^{-7}$ year$^{-1}$ for OGLE-BLAP-009, derived from period-change analysis using OGLE data.

\begin{figure*}
\center\includegraphics[scale=0.46]{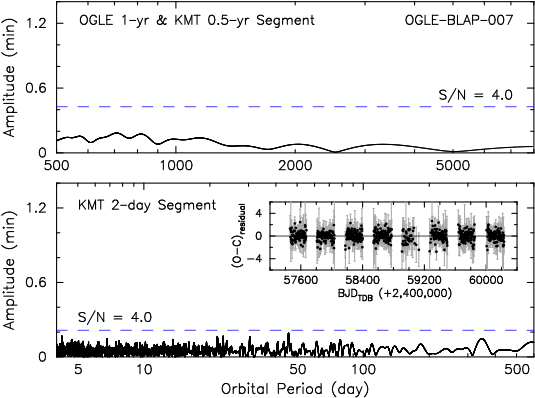} \hskip 3mm \includegraphics[scale=0.46]{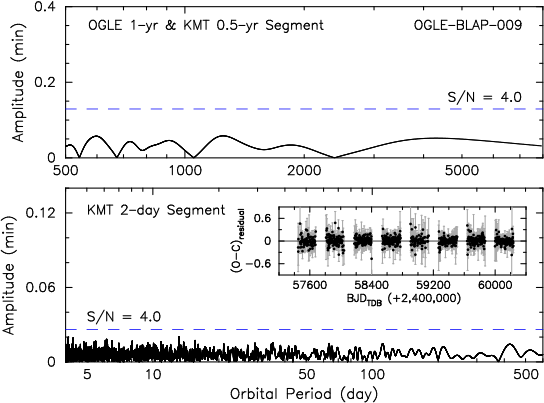}
\caption{Fourier spectra of the $O-C$ residuals for OGLE-BLAP-007 (left) and OGLE-BLAP-009 (right). The other descriptions match those in Figure \ref{fig_blap06_ocres}. \label{fig_blap0709_ocres}}
\end{figure*}

We analyzed the $O-C$ residuals for both BLAPs using the Fourier method to check for periodic signals. The Fourier spectra in the upper panels of Figure \ref{fig_blap0709_ocres} do not exhibit significant peaks in the orbital period range between 500 and 8,000 days, indicating the absence of LTT variations caused by a long-period companion. In addition, we investigated short-period LTT variations using KMTNet $I$-band data, following our analysis method for OGLE-BLAP-006. The lower panels displaying the Fourier spectra of the $O-C$ residuals (inset) for the 2.0-day segmented data show no significant peaks in the orbital period range between 4 and 600 days. Consequently, we detected no LTT-like periodic signal in the pulsation timing variations of OGLE-BLAP-007 and OGLE-BLAP-009, suggesting a low probability of stellar companions around these two BLAPs.

\section{Summary and Discussion \label{sec_discussion}}
This paper presents the first results of our systematic survey to search for stellar companions orbiting BLAPs. We provide long-term, high-cadence photometric data from KMTNet for three BLAPs in the Galactic bulge: OGLE-BLAP-006, OGLE-BLAP-007, and OGLE-BLAP-009. The data collected approximately every 15 minutes over eight consecutive years are invaluable for investigating their pulsation properties and timing variations. To extend the coverage of timing variations further, we analyzed archival data from OGLE.

Our Fourier analysis of the KMTNet $I$-band data detected two new independent frequencies $f_2$ = 38.25 day$^{-1}$ and $f_5$ = 35.05 day$^{-1}$ for OGLE-BLAP-006 alongside the dominant pulsation frequency $f_1$ = 37.88 day$^{-1}$ (see Table \ref{tab_blap06_freq}). Assuming $f_1$ corresponds to the fundamental radial mode, the frequency ratios $f_1 / f_2$ = 0.99 and $f_1 / f_5$ = 1.08 indicate $f_2$ and $f_5$ as nonradial modes. These ratios deviate significantly from the theoretically predicted range of 0.71--0.81 for the ratio between the fundamental and first overtone radial modes \citep{jadlovsky2024}. Our results position OGLE-BLAP-006 as the fifth BLAP to exhibit multiple pulsation modes, following SMSS-BLAP-1, OGLE-BLAP-001, OGLE-BLAP-030, and ZGP-BLAP-08 \citep{chang2024, pietrukowicz2024, pigulski2024}. These stars are excellent targets for studying the physical characteristics of BLAPs through asteroseismic analysis. In contrast, OGLE-BLAP-007 and OGLE-BLAP-009 were confirmed as single-mode pulsators.

Of the more than 100 known BLAPs, only two have been confirmed to have companion stars: HD 133729 \citep{pigulski2022} and TMTS-BLAP-1 \citep{lin2023}. Following the same approach used for these two stars, this study investigated the pulsation timing variations of the three BLAP targets by analyzing $O-C$ diagrams. Timing variations in OGLE-BLAP-006 cannot be interpreted as a single parabola and require two additional long-period LTT terms, suggesting the presence of companion stars in wide orbits. By contrast, the timing variations of OGLE-BLAP-007 and OGLE-BLAP-009 fit excellently into a single parabola with no evidence of LTT variations. After subtracting pulsation-induced light variations, we examined photometric variability due to close companions such as eclipses, reflection effects, or ellipsoidal modulations. No such remaining variability was detected in our three BLAPs, indicating that the presence of close companions is unlikely.

We propose that OGLE-BLAP-006 is the third BLAP with stellar companions, following HD 133729 and TMTS-BLAP-1. The companion stars exhibit distinct physical properties. HD 133729 has a more massive companion in a close, circular orbit with an orbital period $P_{\rm orb}$ = 23.0843 $\pm$ 0.0002 days, eccentricity $e$ = 0.006 $\pm$ 0.002, and mass function $f(M)$ = 1.424 $\pm$ 0.004 $M_\sun$ \citep{pigulski2022}. By contrast, TMTS-BLAP-1 has a less massive companion in a wide, eccentric orbit with $P_{\rm orb}$ = 1,576 $\pm$ 18 days, $e$ = 0.53 $\pm$ 0.05, and $f(M)$ = 0.0012 $\pm$ 0.0001 $M_\sun$ \citep{lin2023}. OGLE-BLAP-006 has two companions with properties similar to those of TMTS-BLAP-1's companion, but with much longer orbital periods: $P_{\rm orb}$ = 4,737 $\pm$ 137 days, $e$ = 0.48 $\pm$ 0.05, and $f(M)$ = 0.0051 $\pm$ 0.0011 $M_\sun$ for the inner companion, and $P_{\rm orb}$ = 6,320 $\pm$ 301 days, $e$ = 0.63 $\pm$ 0.07, and $f(M)$ = 0.0024 $\pm$ 0.0010 $M_\sun$ for the outer one.

\begin{figure}
\center\includegraphics[scale=0.45]{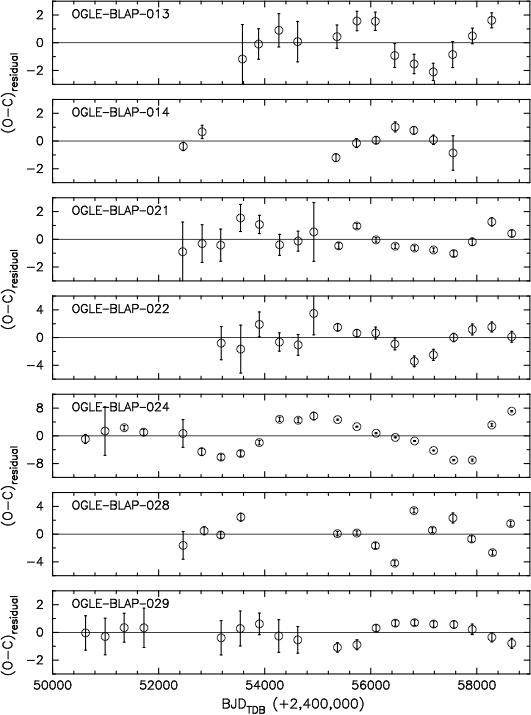}
\caption{$O-C$ residuals (in minutes) for seven BLAP candidates with wide-orbit companions, derived from OGLE data segmented by year. The plot shows periodic variations likely due to the LTT effect, particularly evident for OGLE-BLAP-024. \label{fig_ogle_candidates}}
\end{figure}

It was theoretically unexpected for BLAPs to have companion stars with orbital periods over several thousand days (see Section \ref{sec_intro}). The unexpected detection of these companions led us to investigate other BLAPs. Using OGLE data, we examined the pulsation timing variations of BLAPs discovered in the inner Galactic bulge. Of the 36 BLAPs listed in \citet{pietrukowicz2024}, three from this study and eight with insufficient data were excluded from the analysis. Although this preliminary analysis requires confirmation from additional data, such as that from KMTNet, we identified seven BLAP candidates that exhibited long-period LTT variations. Figure \ref{fig_ogle_candidates} shows the $O-C$ residuals obtained by subtracting the fitted parabola. Recently, \citet{pigulski2024} presented $O-C$ diagrams for OGLE-BLAP-001 and ZGP-BLAP-08, both located in the Galactic disk, revealing long-term periodic variations probably due to the LTT effect. Therefore, BLAPs with stellar companions in wide orbits are likely relatively common. Among the seven BLAPs listed in \citet{pietrukowicz2024} with a negative rate of change in the pulsation periods, six except OGLE-BLAP-007 appear to have such companions. The high fraction of wide-orbit companions in these period-decreasing BLAPs may offer valuable insights into their formation channels.

A plausible scenario is that a BLAP with a wide-orbit companion may be the merger remnant of an inner, close binary in a hierarchical triple system. A solar-type main-sequence binary with a short orbital period of a few days will evolve through mass transfer into an EL CVn-type binary with an ELM pre-white dwarf \citep{chen2017}. The EL CVn binary will then transition to a system similar to TYC 6992-827-1, consisting of an ELM white dwarf and a subgiant, eventually merging into a single white dwarf \citep{lagos2024}. Most of these close binaries are found to have tertiary companions in wide orbits \citep{tokovinin2006, lagos2020, lagos2024, lee2024}. Such hierarchical triple systems will transform into wide binaries by merging the inner close binary, with some merger remnants likely evolving into BLAPs. Therefore, our discovery of wide-orbit companions supports the formation of BLAPs via binary merger proposed by \citet{zhang2023} and \citet{kolaczek2024}.

Future work will investigate the seven identified BLAP candidates with wide-orbit companions using KMTNet data.

\begin{acknowledgements}
This research has made use of the KMTNet system operated by the Korea Astronomy and Space Science Institute (KASI) at three host sites of CTIO in Chile, SAAO in South Africa, and SSO in Australia. Data transfer from the host site to KASI was supported by the Korea Research Environment Open NETwork (KREONET). This research was supported by KASI under the R\&D program (Project No. 2025-1-830-05) supervised by the Ministry of Science and ICT. We thank Editage (www.editage.com) for English language editing.
\end{acknowledgements}

\facility{KMTNet, OGLE.}
\software{{\tt emcee} \citep{foremanmackey2013}, {\tt matplotlib} \citep{hunter2007}, {\tt numpy} \citep{harris2020}, {\tt pyDIA} \citep{albrow2017}, {\tt scipy} \citep{virtanen2020}.}

\section*{Appendix}
The observed times of maximum light, epochs, and $O-C$ values are listed in Table \ref{tab_timing}. These timings derive from the primary frequencies for the three BLAPs:  37.8797546 day$^{-1}$ for OGLE-BLAP-006, 40.9305567 day$^{-1}$ for OGLE-BLAP-007, and 45.0911785 day$^{-1}$ for OGLE-BLAP-009.
 
\begin{deluxetable*}{lrrclrrclrr}
\tabletypesize{\footnotesize}
\tablecaption{Timings for the primary frequencies of three BLAPs. \label{tab_timing}}
\tablehead{ 
   \multicolumn{3}{c}{OGLE-BLAP-006} & & \multicolumn{3}{c}{OGLE-BLAP-007} & & \multicolumn{3}{c}{OGLE-BLAP-009} \\
   \cline{1-3} \cline{5-7} \cline{9-11} 
   \colhead{Times (BJD$_{\rm TDB}$)$^\dagger$} & \colhead{Epoch} & \colhead{$O-C^\dagger$} & &
   \colhead{Times (BJD$_{\rm TDB}$)$^\dagger$} & \colhead{Epoch} & \colhead{$O-C^\dagger$} & &
   \colhead{Times (BJD$_{\rm TDB}$)$^\dagger$} & \colhead{Epoch} & \colhead{$O-C^\dagger$} }
\startdata
OGLE data \\
2,450,614.08997(56)   & $-$261,361 & $-$38.97 & & $\;\;\;\;\;\;\;\;\;\;\;\;\;\;$ -  &    - $\;\;\;\;$  &     - $\;\;$ & & $\;\;\;\;\;\;\;\;\;\;\;\;\;\;$ - &   - $\;\;\;\;$  &  - $\;\;$ \\
2,451,004.43337(94)   & $-$246,575 & $-$34.70 & & $\;\;\;\;\;\;\;\;\;\;\;\;\;\;$ -  &    - $\;\;\;\;$  &     - $\;\;$ & & $\;\;\;\;\;\;\;\;\;\;\;\;\;\;$ - &   - $\;\;\;\;$  &  - $\;\;$ \\
2,451,348.18381(108) & $-$233,554 & $-$27.77 & & $\;\;\;\;\;\;\;\;\;\;\;\;\;\;$ -  &    - $\;\;\;\;$  &     - $\;\;$ & & $\;\;\;\;\;\;\;\;\;\;\;\;\;\;$ - &   - $\;\;\;\;$  &  - $\;\;$ \\
2,451,736.09910(53)   & $-$218,860 & $-$22.60 & & $\;\;\;\;\;\;\;\;\;\;\;\;\;\;$ -  &    - $\;\;\;\;$  &     - $\;\;$ & & $\;\;\;\;\;\;\;\;\;\;\;\;\;\;$ - &   - $\;\;\;\;$  &  - $\;\;$ \\
2,452,077.13018(76)   & $-$205,942 & $-$16.00 & & $\;\;\;\;\;\;\;\;\;\;\;\;\;\;$ -  &    - $\;\;\;\;$  &     - $\;\;$ & & $\;\;\;\;\;\;\;\;\;\;\;\;\;\;$ - &   - $\;\;\;\;$  &  - $\;\;$ \\
2,452,474.23311(221) & $-$190,900 &   $-$9.86 & & $\;\;\;\;\;\;\;\;\;\;\;\;\;\;$ -  &    - $\;\;\;\;$  &     - $\;\;$ & & 2,452,463.15928(14) & $-$227,758 &  +11.43 \\ 
2,452,849.20776(111) & $-$176,696 & $-$11.85 & & 	2,452,823.59229(101) & $-$191,980 & $-$19.90 & & 2,452,823.93818(18) & $-$211,490 &    +9.70 \\
2,453,165.05013(113) & $-$164,732 & $-$10.65 & & 2,453,177.38849(85)   & $-$177,499 & $-$17.25 & & 2,453,174.84888(10) & $-$195,667 &    +8.97 \\ 
2,453,541.66532(115) & $-$150,466 &   $-$7.19 & & 2,453,552.04984(110) & $-$162,164 & $-$13.81 & & 2,453,543.81158(13) & $-$179,030 &    +7.78 \\
2,453,857.21644(44)   & $-$138,513 &   $-$7.23 & & 2,453,899.07670(51)   & $-$147,960 & $-$13.72 & &	 2,453,895.09905(8)  & $-$163,190 &    +6.70 \\ 
2,454,240.11440(86)   & $-$124,009 &   $-$4.16 & & 2,454,254.63241(42)   & $-$133,407 & $-$10.45 & & 2,454,265.68079(7)  & $-$146,480 &    +5.66 \\ 
2,454,618.10260(76)   & $-$109,691 &   $-$0.35 & & 2,454,627.75309(67)   & $-$118,135 &   $-$9.14 & & 2,454,626.34939(6)  & $-$130,217 &    +4.77 \\
2,454,924.28236(265) &   $-$98,093 &      +0.19 & & $\;\;\;\;\;\;\;\;\;\;\;\;\;\;$ -  &    - $\;\;\;\;$  &     - $\;\;$ & & 2,454,922.32715(22) & $-$116,871 &    +4.32 \\
2,455,402.08251(64)   &   $-$79,994 &   $-$1.62 & &	2,455,375.41195(22)    &  $-$87,533 &   $-$5.85 & & 2,455,384.36789(32) &  $-$96,037 &    +3.09 \\
2,455,739.62394(28)   &   $-$67,208 &   $-$2.13 & &	2,455,736.34156(24)    &  $-$72,760 &   $-$4.10 & & 2,455,716.96019(10) &  $-$81,040 &    +2.41 \\ 
2,456,100.10760(24)   &   $-$53,553 &   $-$0.89 & & 2,456,099.37163(21)    &  $-$57,901 &   $-$3.30 & & 2,456,082.95159(14) &  $-$64,537 &    +1.89 \\ 
2,456,468.30007(32)   &   $-$39,606 &      +0.64 & & 2,456,462.08439(21)    &  $-$43,055 &   $-$2.06 & & 2,456,447.90040(10) &  $-$48,081 &    +0.99 \\  
2,456,827.36060(26)   &   $-$26,005 &      +5.38 & & 2,456,825.99414(22)    &  $-$28,160 &   $-$1.05 & & 2,456,810.76495(8)   &  $-$31,719 &    +0.68 \\ 
2,457,185.35901(26)   &   $-$12,444 &      +1.28 & & 2,457,184.28440(21)    &  $-$13,495 &   $-$0.35 & & $\;\;\;\;\;\;\;\;\;\;\;\;\;\;$ - &   - $\;\;\;\;$  &  - $\;\;$ \\
2,457,563.89747(64)   &        +1,895 &   $-$0.85 & & 2,457,563.87910(28)    &       +2,042 &      +0.44 & & 2,457,525.98202(8)   &          +531 &    +0.06 \\ 
2,457,912.28698(47)   &      +15,092 &   $-$4.31 & & 2,457,913.91115(36)     &     +16,369 &      +0.67 & & 2,457,914.99356(4)   &     +18,072 & $-$0.30 \\ 
2,458,281.34485(50)   &      +29,072 & $-$11.09 & & 2,458,282.29125(33)     &     +31,447 &      +0.76 & & 2,458,285.75334(6)   &     +34,790 & $-$0.45 \\ 
2,458,657.40358(38)   &      +43,317 & $-$10.62 & & 2,458,658.26959(40)     &     +46,836 &      +0.86 & & 2,458,635.15644(7)   &     +50,545 & $-$0.52 \\ 
KMTNet data \\													
2,457,513.87134(11)   &                 0 &         0.00 & & 2,457,513.98942(13)     &                0 &        0.00 & & 2,457,514.20584(1)   &                0 &      0.00 \\ 
2,457,614.32016(13)   &        +3,805 &   $-$0.89 & & 2,457,613.08440(14)     &       +4,056 &      +0.45 & & 2,457,614.02577(1)   &       +4,501 & $-$0.05 \\ 
2,457,878.15247(10)   &      +13,799 &   $-$4.57 & & 2,457,877.99663(11)     &     +14,899 &      +0.62 & & 2,457,878.86682(1)   &     +16,443 & $-$0.21 \\ 
2,457,976.01366(10)   &      +17,506 &   $-$6.17 & & 2,457,975.82094(13)     &     +18,903 &      +0.75 & & 2,457,975.44886(1)   &     +20,798 & $-$0.28 \\ 
2,458,242.90797(8)     &      +27,616 & $-$10.32 & & 2,458,243.00543(11)     &     +29,839 &      +1.10 & & 2,458,242.68509(1)   &     +32,848 & $-$0.39 \\ 
2,458,339.58186(10)   &      +31,278 & $-$10.95 & & 2,458,339.26606(11)     &     +33,779 &      +1.14 & & 2,458,339.53329(1)   &     +37,215 & $-$0.40 \\ 
2,458,605.15914(10)   &      +41,338 & $-$10.87 & & 2,458,605.10664(11)     &     +44,660 &      +1.24 & & 2,458,605.19495(1)   &     +49,194 & $-$0.49 \\ 
2,458,707.16628(11)   &      +45,202 & $-$10.66 & & 2,458,708.50129(13)     &     +48,892 &      +1.26 & & 2,458,708.82941(1)   &     +53,867 & $-$0.49 \\ 
2,458,960.73210(17)   &      +54,807 & $-$10.24 & & 2,458,960.61105(19)     &     +59,211 &      +1.01 & & 2,458,960.69685(1)   &     +65,224 & $-$0.50 \\ 
2,459,078.18293(28)   &      +59,256 &   $-$9.92 & & 2,459,082.84262(38)     &    +64,214 &      +1.23 & & 2,459,078.39170(3)   &      +70,531 & $-$0.52 \\ 
2,459,338.79787(10)   &      +69,128 &   $-$8.78 & & 2,459,339.10593(11)      &    +74,703 &     +1.22 & & 2,459,338.26518(1)   &      +82,249 & $-$0.47 \\ 
2,459,437.87493(10)   &      +72,881 &   $-$8.22 & & 2,459,437.07657(13)     &     +78,713 &     +0.97 & & 2,459,438.44002(1)   &      +86,766 & $-$0.42 \\ 
2,459,703.40062(13)   &      +82,939 &   $-$6.39 & & 2,459,702.28176(13)     &     +89,568 &     +0.83 & & 2,459,702.19456(1)   &      +98,659 & $-$0.32 \\ 
2,459,799.83772(14)   &      +86,592 &   $-$5.87 & & 2,459,800.66782(14)     &     +93,595 &     +0.70 & & 2,459,802.48029(1)   &    +103,181 & $-$0.27 \\ 
2,460,067.05334(14)   &      +96,714 &   $-$3.51 & & 2,460,067.85196(13)     &   +104,531 &     +0.55 & & 2,460,068.43041(1)   &    +115,173 & $-$0.13 \\
2,460,167.63515(11)   &    +100,524 &   $-$2.97 & & 2,460,166.79980(13)     &   +108,581 &     +0.20 & & 2,460,168.42784(1)   &    +119,682 & $-$0.06 \\ 
\enddata
\tablenotetext{^\dagger}{The values in parentheses represent errors in the last digit. $O-C$ values are in minutes.}
\end{deluxetable*}

\end{document}